# 3D Computational Cannula Fluorescence Microscopy enabled by Artificial Neural Networks

RUIPENG GUO,[1] ZHIMENG PAN,[2] ANDREW TAIBI,[3] JASON SHEPHERD,[3] RAJESH MENON[1,*]

[1] *Department of Electrical and Computer Engineering, University of Utah, Salt Lake City, UT 84112, USA.*
[2] *School of Computing, University of Utah, Salt Lake City, UT 84112, USA.*
[3] *Department of Neurobiology & Anatomy, Biochemistry and Ophthalmology & Visual Sciences, University of Utah, Salt Lake City, UT 84112, USA.*
*\* rmenon@eng.utah.edu*

**Abstract**: Computational Cannula Microscopy (CCM) is a high-resolution widefield fluorescence imaging approach deep inside tissue, which is minimally invasive. Rather than using conventional lenses, a surgical cannula acts as a lightpipe for both excitation and fluorescence emission, where computational methods are used for image visualization. Here, we enhance CCM with artificial neural networks to enable 3D imaging of cultured neurons and fluorescent beads, the latter inside a volumetric phantom. We experimentally demonstrate transverse resolution of ~6μm, field of view ~200μm and axial sectioning of ~50μm for depths down to ~700μm, all achieved with computation time of ~3ms/frame on a laptop computer.

## 1. Introduction

Neurons are distributed in the brain in a complex manner in 3D space. Therefore, neural imaging requires data in all 3 spatial dimensions, ideally with fast acquisition. Computational Cannula Microscopy (CCM) has shown to be effective for imaging fluorescence deep inside the brain with minimal trauma [1,2]. Previous work also demonstrated the potential of computational refocusing to achieve quasi-3D imaging (in air) [3]. Recently, we also demonstrated the application of Artificial Neural Networks (ANNs) to drastically improve the speed of image reconstructions in CCM [4]. Here, we extend our previous work to enable ANNs to perform fluorescence imaging in a 3D volume. Specifically, we investigated three different ANNs to enable 3D CCM, and demonstrated 3D imaging using cultured neurons and fluorescent beads, both in air and inside a volumetric phantom.

   3D imaging of neurons *in vivo* in the intact brain is typically achieved with 2-photon imaging. Although impressive resolution, field of view and speeds have been demonstrated recently [5], such mesoscopes require fairly expensive equipment, complex procedures and are typically limited to depths of a few hundred micrometers. Many other approaches exist for imaging fixed/dead neurons including clearing tissue to render them transparent [6] and utilizing polymeric expansion techniques [7]. Lightsheet microscopy [8,9] and structured illumination approaches [10] have also been successfully employed for fast high-resolution volumetric imaging of transparent samples. Tomography from multiple 2D images using deep convolutional ANNs have been demonstrated for semiconductor-device inspection using reflected light [11]. Alternative machine-learning approaches have been combined with tomography [12] and light-field microscopy [13]. Computational refocusing can also achieve 3D imaging either with optics-free setups [14-16] or with diffractive masks [17]. Recent work has applied similar principles to 3D wide-field fluorescence microscopy of clear samples using miniaturized mesoscopes [18]. In contrast to these approaches, CCM is able to image deep inside opaque or highly-scattering tissue such as the mouse brain [2,19]. Furthermore, CCM has the advantage that the ratio of field-of-view to probe diameter is close to 1, thereby allowing for minimal invasive imaging. In our experiments, the probe (cannula) diameter is 220μm. Alternative approaches to imaging through multi-mode fibers have also been described,[20-23] but most of these rely on the temporal coherence of the excitation light, and thereby require more complex equipment and computational methods.

## 2. Experiment

The schematic of our CCM setup is shown in Fig. 1a (and Fig. S1) [24]. The cannula (FT200EMT, Thorlabs) can be inserted into the sample. The excitation light (LED with center wavelength = 470 nm, M470L3, Thorlabs) is coupled to the cannula through an objective lens. Fluorescence from the sample is collected by the same cannula (therefore, an epi-configuration). The image at the top (distal) end of the cannula is imaged onto a sCMOS camera (C11440, Hamamatsu). Reflected excitation light is rejected by a dichroic mirror and an additional filter. An exemplary image is shown on the right inset. A reference microscope is placed underneath the sample to image the same region as the cannula and the corresponding image is shown in the right inset as well. First, we performed a series of experiments to determine that the volume of interest (constrained by the fluorescence collection and excitation efficiencies) is limited to approximately 100μm from the bottom surface of the cannula. Subsequently, we restricted our experiments to 3 layers spaced by 50μm in close proximity to the cannula (as illustrated in the left inset in Fig. 1a).

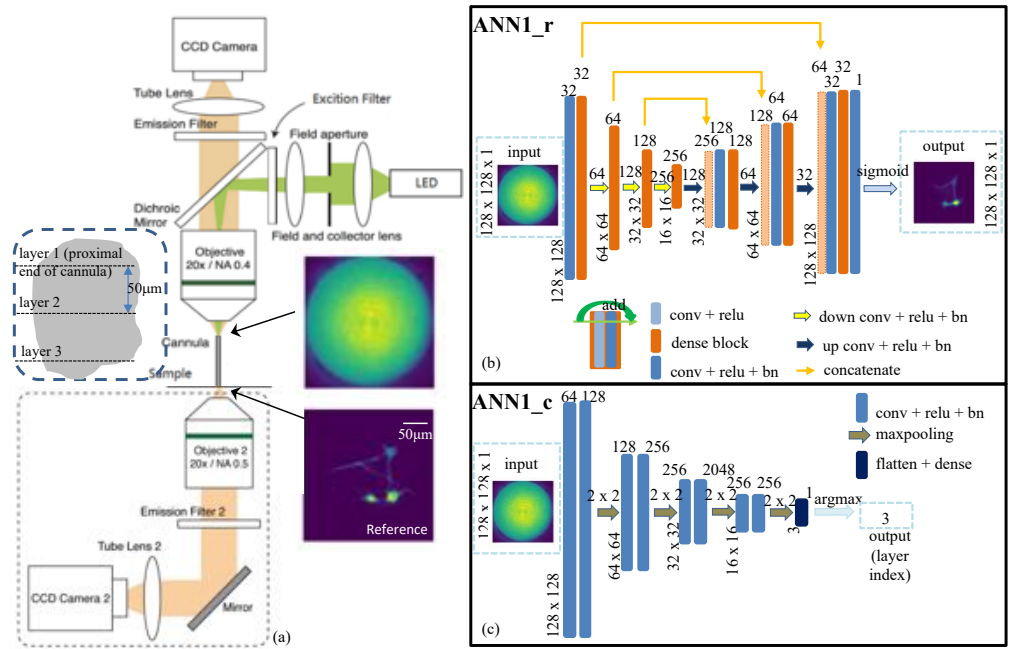

**Fig. 1.** Overview of computational-cannula microscopy (CCM). (a) Schematic of microscope. Left inset shows 3 layers that are captured, spaced by 50μm from the proximal end of the cannula. Right insets show recorded images with CCM (top) and with the reference microscope (bottom). (b) Details of ANN1_r that is trained to take the input CCM image and output the reconstructed image of 1 layer. A modified version of this network, ANN2 outputs 3 images, one for each layer (see SI) [24]. (c) Details of ANN1_c, which classifies the input CCM image into one of the 3 layers.

Similar to our previous work, [4] here we used both mouse primary hippocampal cultured neurons and slides with fluorescent beads to create a dataset for training the ANNs. However, unlike our previous work, we acquired this dataset for 3 layers as illustrated in the left inset of

Fig. 1a. Details of sample preparation are described in section 2 of the supplement [24]. A total of 16,700 images from each layer were recorded.

Figure 1b shows the architecture of ANN1_r, which is used to convert the input CCM image into the fluorescence image. It consists of dense blocks that prevent the gradients from vanishing too fast. Each dense block includes 3 individual layers: 2 convolutional layers with RELU activation function followed by a batch-normalization layer. The structure is a typical U-net with the skip connections to concatenate the encoder and decoder outputs [4]. A second ANN, referred to as ANN1_c (Fig. 1c) was used to classify the images into the 3 layers (layer index is metadata in the images). It includes 8 blocks and a final classifier. Each block consists of one 2D convolution layer with RELU activation function followed by a batch-normalization layer. We add a Maxpool function between every two blocks, which down-samples input images and prevents overfitting by taking max value in a filter region (2x2 filter). For both ANNs, the loss function is the pixel-wise cross-entropy, defined as:

$$L = \frac{1}{N}\sum_i -g_i \log p_i - (1 - g_i)\log(1 - p_i),$$

where $g_i$ and $p_i$ represent the ground truth and predicted pixel intensity, respectively. This loss function imposes sparsity. An alternative ANN, referred to as ANN2 that outputs 3 layer images was also explored and described in section 3 of the supplement [24].

## 3. Results

Imaging results using ANN1_r and ANN1_c for cultured neurons and fluorescent beads are summarized in Fig. 2a. The ground-truth images were obtained by the reference microscope as described earlier and confirm the accuracy of the ANN outputs. The layer index predicted by ANN1_c was verified by the metadata of the corresponding reference images. The structural similarity index (SSIM) and maximum-average error (MAE), both averaged over 1000 test images, were 90% and 1% for ANN1_r, respectively. The classification accuracy averaged over 1000 images, of ANN1_c was 99.8%. Figures 2b and 2c show results from ANN2 with cultured neurons and fluorescent beads, respectively.

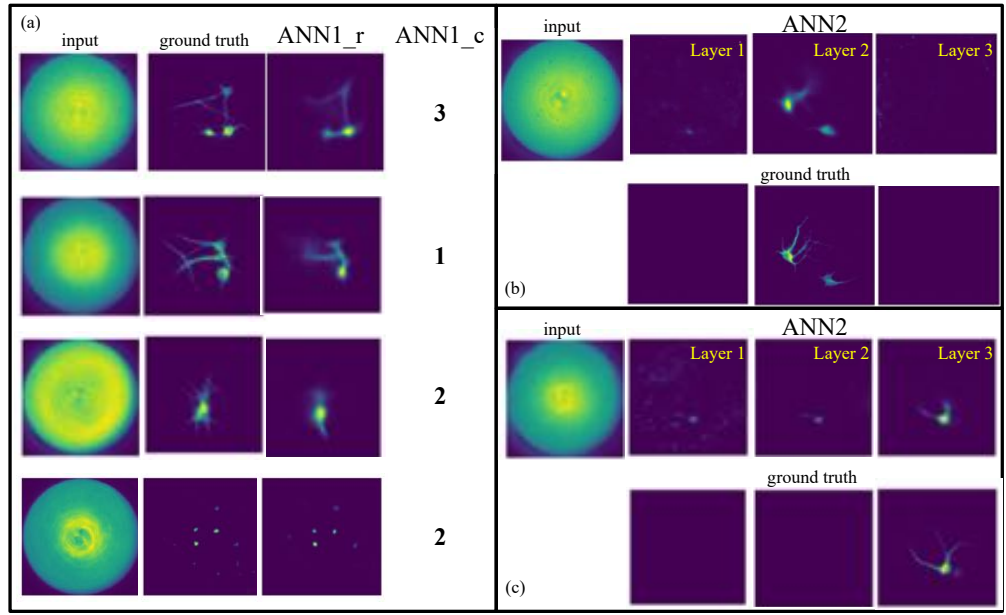

**Fig. 2.** Experimental results of 3D CCM. (a) Fluorescence samples were reconstructed using ANN1_r, while the layer index was predicted by ANN1_c. The first 3 rows show cultured neurons, while the last row shows fluorescent beads (diameter=4μm). (b) ANN2 produces 3 reconstructed images, one for each layer. This is an example, where layer 2 contains the neuron. (c) Another example from ANN2 where layer 3 contains the neuron. Many additional examples from all networks are included in the SI [24].

The performance of ANN2 averaged over 1000 test images and 3 layers per image was 96% (SSIM) and 0.4% (MAE). We further evaluated the computation time on a computer equipped with Intel(R) Core(TM) i7-4790 CPU (clock frequency of 3.60 GHz, memory of 16.0 GB) and NVIDIA GeForce GTX 970. The average reconstruction time for ANN1_r and ANN2 was 3.3ms and 3.4ms, respectively. The average classification time for ANN1_c was 3.6ms.

We also compared the performance of the two reconstruction ANNs by applying these to the same input image in Fig. 3a and ground-truth image in Fig. 3b (layer index is labelled 2). The reconstructed result from ANN1_r is in Fig. 3c. The output of ANN1_c is 2. The corresponding output from ANN2 is shown in Figs. 3d-f for layers 1, 2 and 3, respectively.

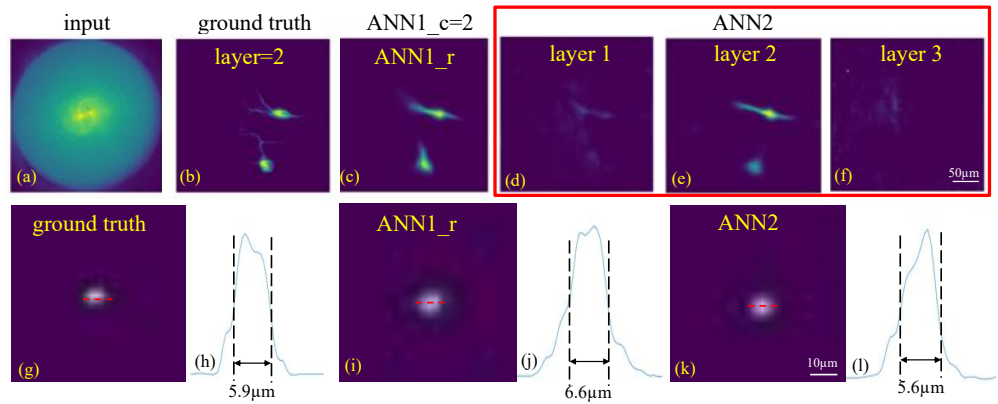

**Fig. 3.** (a-f) Comparison of results from the different reconstruction ANNs using the same input image. In (c), the output of ANN1_c is labeled as 2 on top. (g-k) shows images of a single fluorescent bead (diameter=4μm) obtained using the 2 networks (the bead is in layer 1). Size of images (a)-(f) are the same and those of (g),(i) and (k) are the same.

In order to estimate resolution, we imaged a single fluorescent bead, whose ground truth image and cross-section through the bead are shown in Figs. 3g and 3h, respectively. A bead diameter of 5.9μm is measured. The corresponding outputs from ANN1_r and ANN2 are in Figs. 3i-l, respectively. The corresponding measured bead diameters were 6.6μm and 5.6μm, respectively.

Finally, we fabricated a phantom made of agarose dispersed with fluorescent beads as illustrated by the photographs in Fig. 4a [24]. The cannula was carefully inserted into the phantom, while CCM images were recorded. The ANN1_r was retrained with a synthetic dataset comprised of combining the 3-layer CCM images into a single "synthetic" CCM image (see section 4 of supplement [24]). We refer to this new network as ANN1_r*, which is now trained to reconstruct an image comprised of the projection of the fluorescence signal from within 100μm of the proximal end of the cannula onto a single plane. The CCM images and corresponding output images of ANN1_r* at various depths are shown in Fig. 4b. Only a subset of the images are shown here and the complete set is included in section 5 of the supplement [24]. This stack of 2D images can then be combined into a reconstructed 3D image as shown in Fig. 3c [24].

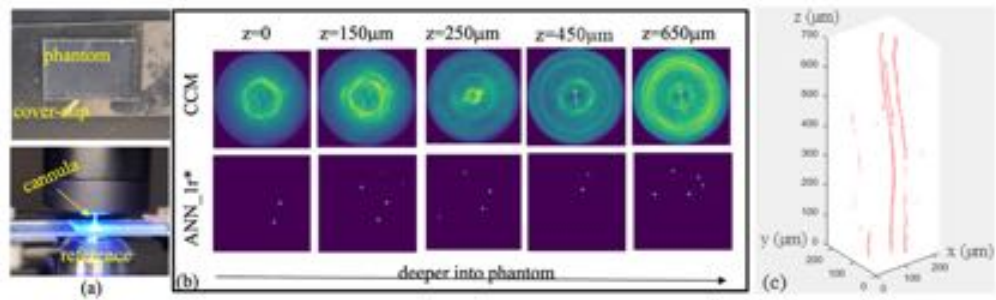

**Fig. 4.** Imaging inside a volumetric phantom: (a) Photographs of phantom. Bottom image shows cannula inserting into the phantom. Blue light is excitation. (b) Output of ANN1_r* trained on a synthetic dataset (see text for details) at various z, depths inside the phantom (see Visualization 1). (c) Reconstructed 3D image. The streaks in the z direction indicate that the cannula pushed some beads down inside the phantom during the experiment (see Visualization 2).

One of the advantages of ANN over previous approaches that utilize singular-value decomposition (SVD) [1-3] is the much higher computation speed. In table 1, we summarize the performance of the 2 ANN approaches and the SVD method. The performance of the 2 ANN approaches are similar, and were averaged over 1000 test images, each containing 3 layers. Classification accuracy was defined as the ratio of the number of images with correctly predicted layer index to the total number of images tested (1000). Data used for SVD was from a single layer data, hence classification accuracy is not applicable.

Table 1: comparison of the performance of each ANN and linear algorithm.

|  | Structural similarity index (SSIM) | Mean absolute error (MAE) | Computation time | Classification accuracy |
|---|---|---|---|---|
| ANN1 | 0.8974 | 0.0104 | (r)3.3ms, (c)3.6ms | 0.9980 |
| ANN2 (3 layer average) | 0.9639 | 0.0036 | 3.4ms | NA |
| SVD | 0.9576 | 0.0138 | 100ms | NA |

## 4. Conclusion

In conclusion, we demonstrate 3D imaging of fluorescent beads in a volumetric phantom using a surgical cannula as the lightpipe for both excitation and fluorescence. Image reconstructions in multiple planes were achieved using trained artificial neural networks. We trained two types of neural networks on both experimental and quasi-experimental data (augmented by synthesizing multiple plane images together). The system was able to achieve lateral resolution of ~6μm, axial sectioning of ~50μm and imaging depths as large as 0.7mm. The field of view was approximately equal to the diameter of the cannula, thereby allowing for imaging a wide area with minimal invasive surgery.

**Funding**
National Science Foundation BRAIN Grant # 1533611
National Institutes of Health R21 Grant# 1R21EY030717

**Disclosures**
The authors declare no conflicts of interest.

See Supplement 1 for supporting content.

# 3D Computational Cannula Fluorescence Microscopy enabled by Artificial Neural Networks: Supplementary Information

RUIPENG GUO,[1] ZHIMENG PAN,[2] ANDREW TAIBI,[3] JASON SHEPERD,[3] RAJESH MENON[1],*

[1]Department of Electrical and Computer Engineering, University of Utah, Salt Lake City, UT 84112, USA.
[2]School of Computing, University of Utah, Salt Lake City, UT 84112, USA.
[3]Department of Neurobiology & Anatomy, Biochemistry & Ophthalmology & Visual Sciences, University of Utah, Salt Lake City, UT 84112, USA.
* rmenon@eng.utah.edu

## 1. CCM Hardware Description.

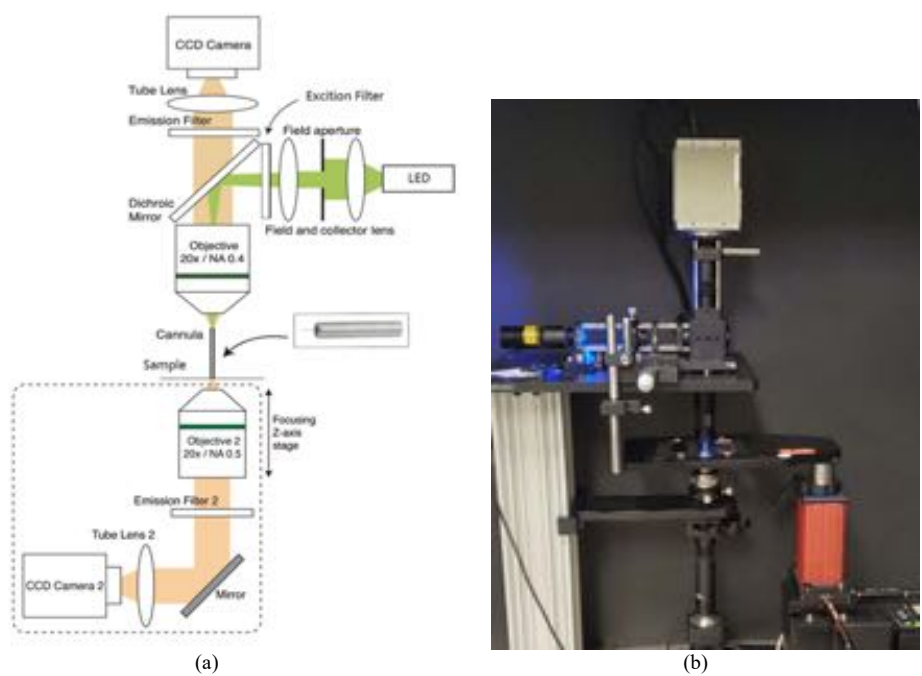

**Fig. S1.** Setup of CCM system: (a) Sketch of CCM system. (b) Photograph of the CCM setup.

The schematic and photograph of our Computational Cannula Microscope (CCM) is shown in Fig. S1. Excitation from a blue LED (center wavelength D 470 nm, M470L3, Thorlabs) is conditioned and focused onto the top face of the cannula via a 20X objective (PLN 20X, Olympus). The cannula guides the excitation to its bottom face and uniformly illuminates a sample placed in close proximity (~100 μm in this paper). The fluorescence from the sample is collected by the same cannula and guided to its top face, which is then imaged onto a sCMOS camera (C11440, HAMAMATSU). We set up a 520 nm–35 nm filter and a 472 nm–30 nm filter in the optical path to separate the fluorescent signals from the source beam. The cannula is made by removing the sheath of the fiber (FT200EMT, Thorlabs) and placing it in a stainless-steel ferrule (SFLC230, Thorlabs). The cannula diameter and the length are 220 µm and 7 mm, respectively. The numerical aperture of the cannula is 0.39, which is close to those of the objectives. A reference fluorescence microscope was equipped to image the same sample from underneath with objectives that have the same magnification as the one in CCM. The field of view (diameter of circle in the CCM image) is 200 µm, and that of the reference microscope is ~260 µm.

## 2. Samples preparation

### 2.1 Fluorescent beads sample

We utilized 4µm green FluoSpheres sulfate microsphere (505/515) in this paper. Spectrum of the micro-bead is shown in the Fig. S2. To make samples with desired density, diluted the bead solution (2% solid) to 1:500 high purity water, then vortex the diluted solution for 2 minutes. The 1:500 solution is diluted once more to 1:10 high purity water to make 1:5,000 solution, followed by vortex. Then aliquot 50uL of 5,000 solution onto a glass slide, until it is completely dry. We do not use coverslip to locate the bead sample as possible as close to the distal end of cannula. Once prepared, we check the distribution of beads on the slide under the reference microscope. The slide should include both sparse beads area and dense beads area to build a good dataset.

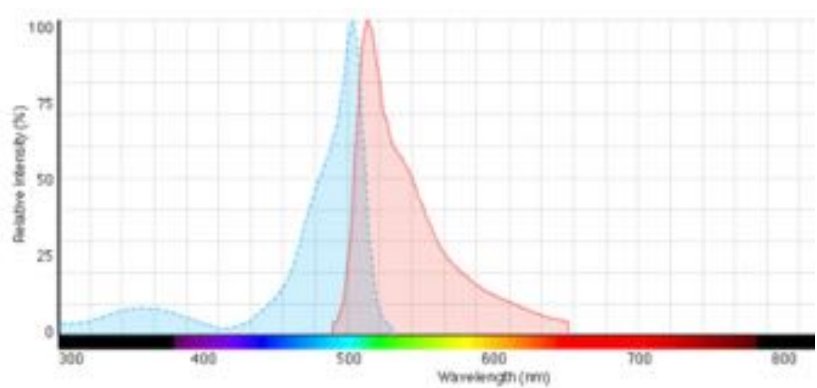

**Fig. S2.** Green FluoSpheres microsphere (diameter is 4µm) is used in the experiment. The blue dash line is excitation curve and the red line is emission curve.

### 2.2 Cultured neurons sample

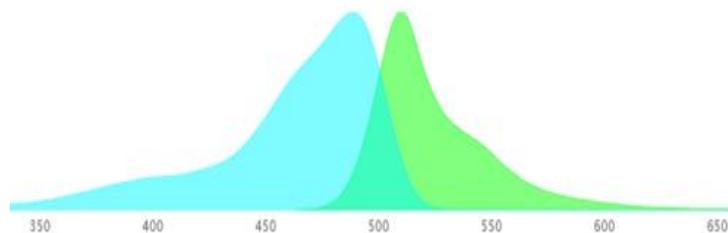

**Fig. S3.** The spectrum of pCAG-eGFP.

Same neurons sample was used as the previous work[1]. Primary neurons were taken from dissociated hippocampi of E18.5 Sprague-Dawley rat pups. Hippocampi were dissociated using 0.01% DNase (Sigma-Aldrich) and 0.067% papain (Worthington Biochemicals) prior to trituration through glass pipettes to obtain a single-cell suspension. Cells were then plated at 8 × 104 cells/ml in the Neurobasal medium (Thermo-Fisher) supplemented with 5% horse serum, 2% GlutaMax (Thermo Fisher), 2% B-27 (Thermo Fisher), and 1% penicillin/streptomycin (Thermo Fisher) on coverslips (No. 1, Bioscience Tools) coated overnight with 0.2 mg/ml poly-L-lysine (Sigma-Aldrich) in 100mM Tris-base (pH 8). Neurons were grown at 37◦C/5% CO and fed via a half-media exchange every third-day with astrocyte-conditioned Neurobasal media supplemented with 1% horse serum, 1% GlutaMax, 2% B-27,and 1% penicillin/streptomycin, with the first feeding containing 5 µM β-D-arabinofuranoside (Sigma-Aldrich) to limit the overgrowth of the glial cells. The neurons were grown for 12–14 days in vitro prior to transfection, fixation, and imaging.

The neurons were transfected after 12 days in vitro with 0.5 µg of pCAG-eGFP (Addgene: 89684) using lipofectamine 2000 at a 3:1 ratio when complexed with plasmid DNA. The spectrum of pCAG-eGFP is shown in Fig. S3. Then neurons were transfected over the course of 1 h at 37◦C in pH 7.4 Minimum Essential Media (Thermo Fisher) supplemented with 2% GlutaMax, 2% B-27, 15 mM HEPES (Thermo Fisher), 1 mM Sodium Pyruvate (Thermo Fisher), and 33 mM Glucose. After the transfection, the neurons were given 24 hrs in growth media at 37◦C/5% CO2 to allow a sufficient recovery and expression of the plasmid prior to fixation in 4% formalde-hyde (thermo fisher)/4% sucrose (VWR) in phosphate buffered saline for 15 min at room temperature. After fixation, the neurons were mounted in a Prolong Gold Aqueous Medium (Thermo Fisher) and imaged.

2.3 Agarose phantom preparation

We uses the similar method with that used in Kim's paper[2]. Imaging phantom used for experiment was made in 2% agarose gel, prepared by the following steps. First, mix 0.1g of agarose powder (EZ BioResearch S-1020-500) and 5mL of high purity water in a flask and stir it until agarose disperses uniformly in the solution. Then heat it in a microwave oven, using 100% power for 10 seconds and repeat heating with 10 seconds interval until agarose completely dissolves. Gently stir between intervals to suspend agarose. Before agarose solidify, mix 50uL of 1:500 diluted bead mixture into the agarose and pour the agarose solution into a mold on a glass slide. When agarose solidify, fix agarose gel on the glass slide with glue. The thickness of phantom we used in this paper were 1-1.5mm and 5mm. Additional results from the agarose phantom are shown in supplementary fig. S8-S10.

**3. Description of ANN2**

We modified the reconstruction ANN by translating output of model from one single image to three images, corresponding to three z-positions, as shown in Fig. S4. To adapt this model, we changed recorded reference images to three images by adding two images with zero values. The sequence of recorded reference images in the 3D array are different, related to the z-position. Once the model is trained, it can output three reconstructed images (including the reconstructed image on one layer) with one CCM image input. More results from both ANN1 and ANN2 are shown in Fig. S5-S7.

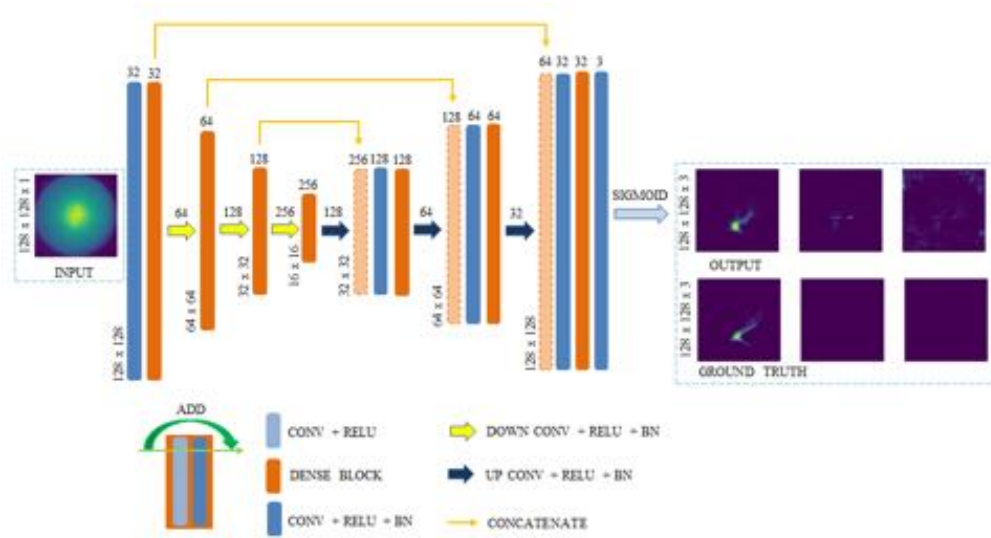

**Fig. S4.** The designed ANN2 system: ANN2 architecture is similar with the reconstruction ANN in ANN systems, but size of output is modified to 128*128*3.

## 4. Dataset and training of ANNs

To build 3D dataset, first we fixed z position and stepped the slide using a stage in a raster fashion with the step size of 50μm. Then we changed the z position of slides by 50μm and the reference microscopy was refocused to record reference image. In this paper, we chose 3 z-positions to build the dataset. Both reference images (1024*1024) and CCM (340*340) images were reshaped to 128*128 to fit the ANN and also to speed up the training process for proof of principle. 16700 images from each layer are used for training and testing the ANN systems. All the images were mixed together and separated into training data, evolution data and testing data. 1000 images were used for testing and 10% percent of other images were used as evolution data. All the images were normalized into range [0,1] before training and testing. The batch size was 16 in all the training of ANNs. And the training began overfitting after about 15 epochs. Once the model was trained, it could be saved for further prediction. For the ANN that used for phantom test, we pre-process the 3 layers dataset with the equation below:

$CCM\_image = CCM\_image(layer1) + CCM\_image(layer2) + CCM\_image(layer3)$
$Ref\_image = Ref\_image(layer1) + Ref\_image(layer2) + Ref\_image(layer3)$

We generated synthetic images that included 3-layers information in both CCM images and reference images with these equations. During the phantom test, first we put the sample under the cannula and moved it close to the end of cannula. Then we moved the sample up with step 50 μm and recorded the CCM images simultaneously. The cannula was inserted into phantom step by step till all the tip of cannula got into the sample. In this test, we can insert cannula into sample as depth as 0.75mm. Note that, this limitation comes from that the length of cannula tip in this test was 0.75mm.

## 5. More Results

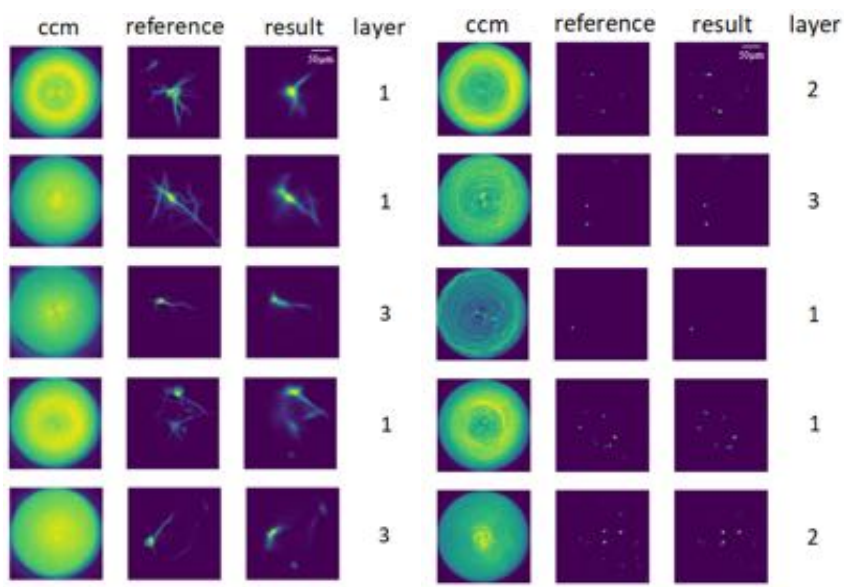

**Fig. S5.** The reconstructed results are from ANN system that consists of reconstruction ANN and classification ANN. (a) The results are from neurons dataset. SSIM and MAE for training dataset are 0.9213 and 0.0082. SSIM and MAE for testing dataset are 0.8974 and 0.0104. The accuracy for classification ANN could reach 0.9980. The layer column is the predicted layer numbers, which correspond to z-positions. (b) Dataset built with beads sample. SSIM and MAE for training dataset are 0.9791 and 0.0017. SSIM and MAE for testing dataset are 0.9878 and 0.0012. The accuracy for classification ANN is 0.9665.

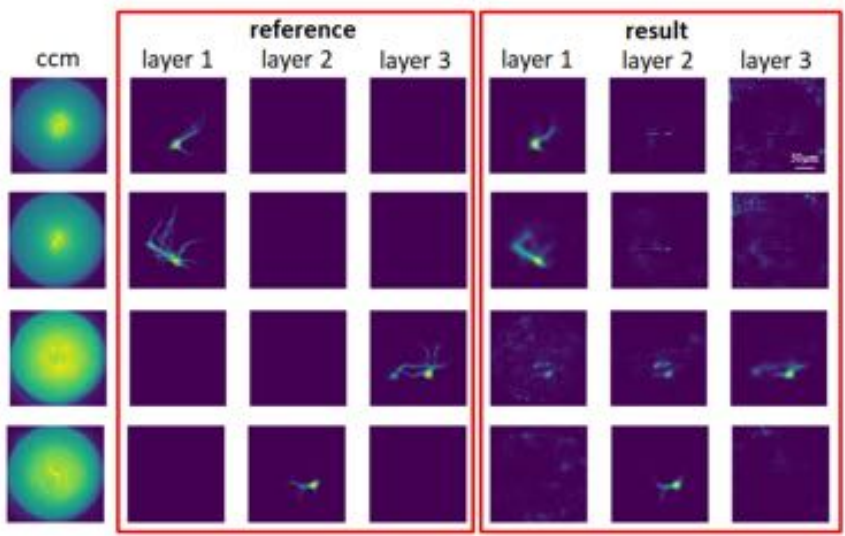

(a)

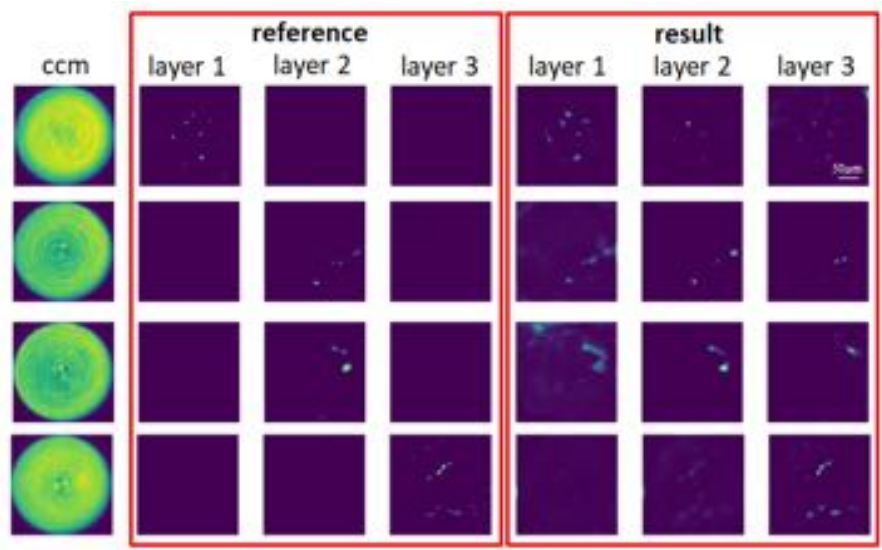

(b)

**Fig. S6.** The reconstructed results are from ANN system that output three images with single ccm image input. The reference images in left red square come from the reference image by adding two images with zero values. The results in right red square are reconstructed from the ccm images in left column. (a) Results come from dataset built with neurons sample. SSIM and MAE for test dataset are 0.9700 and 0.0031. SSIM and MAE for test dataset are 0.9639 and 0.0036. (b) Results come from dataset built with beads sample.

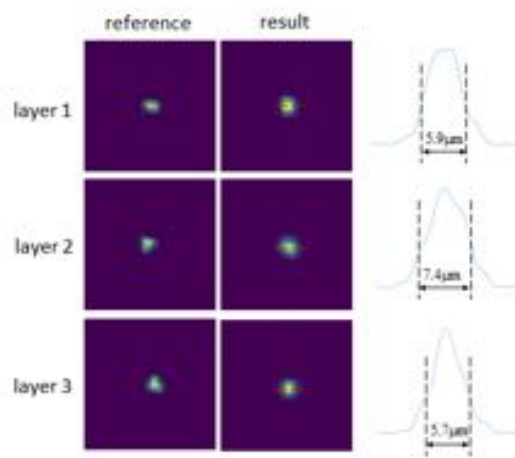

**Fig. S7.** Comparison of reconstructed results in different layers: the results are three single bead images come from the three different layers. The resolution was similar for the three layers.

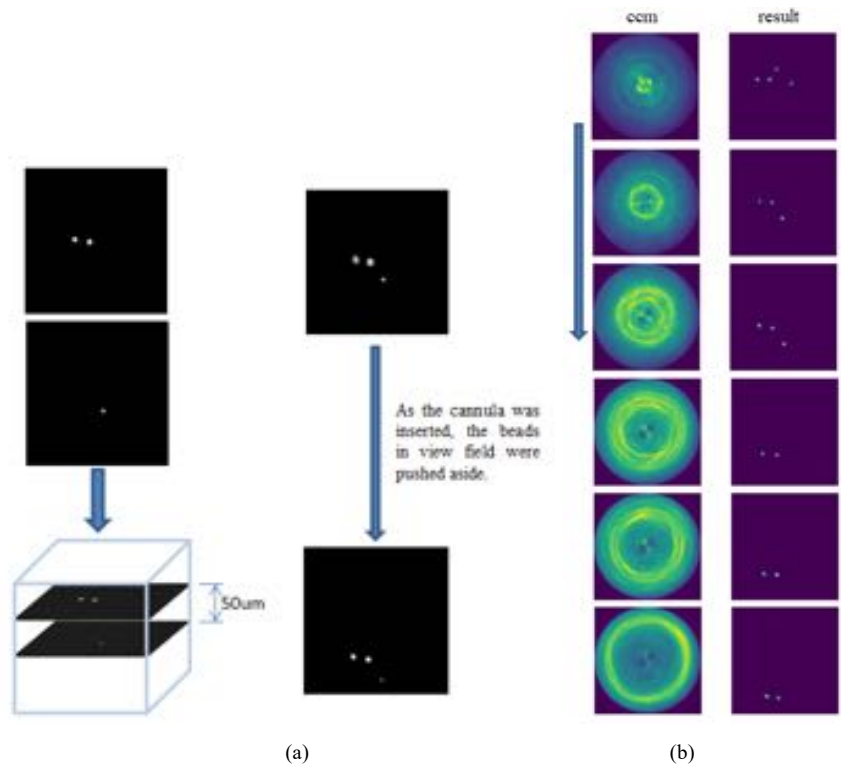

(a)              (b)

**Fig. S8.** More phantom result: the thickness of phantom used in this test is about 1 mm. The beads were sparse in this sample. (a) We can track the beads in view field with reference microscope. Three beads were in view field, two beads were on same z plane and the other one was on different z plane. From the left image, we can see that the beads were pushed aside as the cannula was inserted. (b) The cannula touched sample from the first images. We got similar images from reconstructed results.

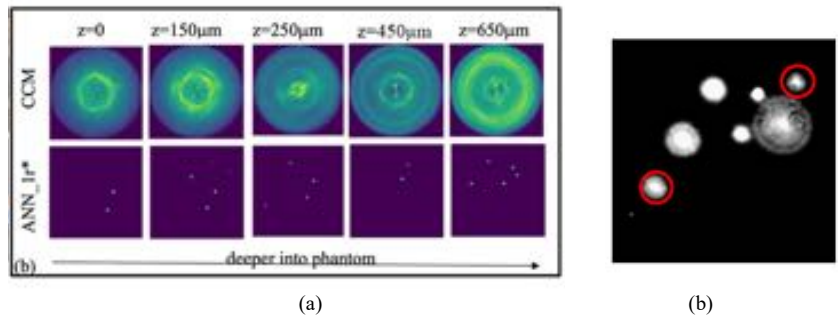

(a)              (b)

**Fig. S9.** For the test shown in fig.4, when the cannula was inserted into phantom as depth as 700μm, we tried to recorded the beads distribution with reference microscope, as shown in (b). We increased the exposure time to make sure all the beads could be seen. Most beads were reconstructed as z = 650μm, but the two aside bead in red circles.

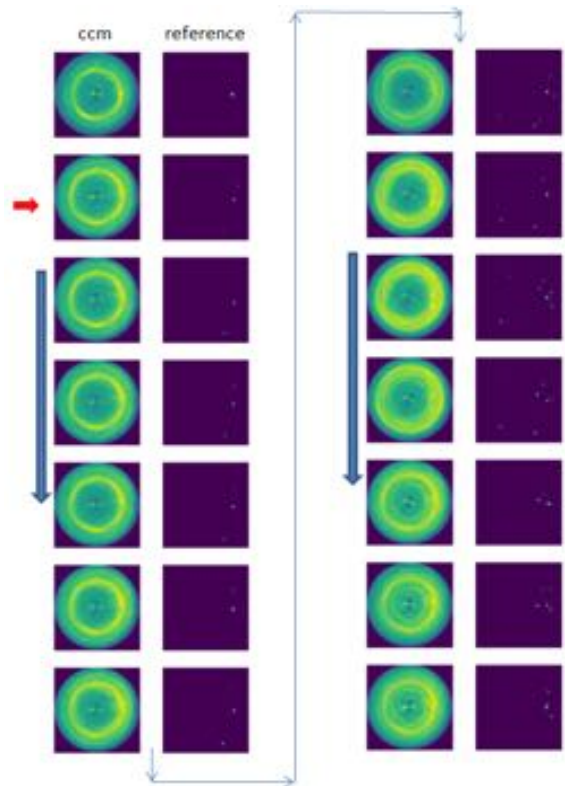

**Fig. S10.** More phantom result: the thickness of phantom used in this test is 5mm. We couldn't track the beads distribution with reference microscope. The cannula began to be inserted into phantom at the red arrows position.